\documentclass{IEEEtran}
\usepackage{cite}
\usepackage{setspace}
\usepackage{cool}
\usepackage{caption}
\usepackage{subcaption}
\usepackage{amsmath,amssymb,amsfonts,xpatch}
\usepackage{amsthm}
\usepackage{lipsum}
\usepackage{algorithmic}
\usepackage{graphicx}
\usepackage{xcolor}
\usepackage[utf8]{inputenc}
\usepackage{amsmath,mathtools}
\usepackage{float}
\author{Farhad Mirkarimi }
\newtheorem{theorem}{Theorem}
\newtheorem{lemma}{Lemma}
\newtheorem{remark}{Remark}

\DeclareMathOperator{\taninv}{tan^{-1}}
\DeclareMathOperator{\erf}{erf}
\allowbreak
\begin{document}
\title{On the Capacity of Joint Time and Concentration Modulation for Molecular Communications}
\author{Farhad Mirkarimi, Mahtab Mirmohseni and Masoumeh Nasiri-Kenari\\
Department of Electrical Engineering, Sharif University of Technology, Tehran, Iran}
\maketitle
\begin{abstract}
Most diffusion based molecular channels suffer from low information capacity due to the structure of the diffusion environment. To address this issue, this paper studies the capacity of the diffusion based molecular communication by exploiting both time and concentration level of the released molecules for information transfer. While the release time can, in general, be any real number, for the sake of tractability, we consider a discrete value for it, by dividing the transmission time interval into some sub-intervals. The transmitter releases molecules in one of the sub-intervals with a level of concentration both determined by input data, hereby applying joint time and concentration (JTAC) modulation. We derive the lower and upper bounds on the JTAC channel capacity. The observation time, at  the receiver, which is equal to symbol period, is divided to some sub-intervals, not necessarily equal to the number of sub-intervals in the transmitter,  and the number of received molecules in each sub-interval is counted. We propose three lower bounds, depending on how the receiver uses the number of molecules counted in the sub-intervals. In the first scheme, the receiver examines the sub-interval which has the maximum mutual information with channel inputs. Second scheme exploits the sum of received molecules in all sub-intervals to determine the concentration level. Then conditioned on the concentration level, for each sub-interval, the mutual information between the release time and the number of molecules received at the sub-interval computed, then the maximum of these mutual information terms determines the release time. In the last scheme,  the concentration is detected using a similar approach to the second scheme, while the difference of received molecules in adjacent sub-intervals is utilized to detect the release time. A closed form lower bound expression has been derived for each case. Moreover, the symmetric Kullback-Liebler (KL) divergence metric is used to obtain a computable upper bound on the channel capacity. Finally, the Blahut-Arimoto algorithm is used to compute the capacity numerically, and to determine how tight the derived bounds are. Our numerical results indicate that the third scheme provides tighter lower bounds on the capacity compared with the two other cases. The improvements compared to the conventional concentration based modulation and timing based modulation are also demonstrated.
\end{abstract}
\section{Introduction}
Molecular communication (MC) is a new paradigm which uses molecules as a carrier for information transfer. It has applications in a wide range areas from under water communications to drug delivery and medicine\cite{DBLP:journals/corr/FarsadYECG14}. Different modulation techniques have been proposed for MC including transmission of information using the concentration, the type, or release time of molecules. In addition, different methods have been used to carry particles from transmitter to receiver such as diffusion, flow, and active transport \cite{DBLP:journals/corr/FarsadMEG16}.
Like any other communication systems, it is important to find limits imposed on this paradigm to prove its applicability and to design efficient transmitter and receiver. With this regard, although significant works have been done on the capacity of MC channels, unfortunately the capacity of most molecular channels are open problem, in particular when realistic conditions are imposed\cite{gohari2016information}. In the early works on the capacity of concentration channels, the authors in\cite{einolghozati2011capacity} have considered a Markov chain model for the channel to capture the memory of the channel, and they have computed the capacity numerically.

In \cite{6648629}, the  quorum sensing phenomena has been considered, where two colonies of bacteria send and receive messages by sensing concentration in each time-slot, and the upper and lower bounds on the channel capacity have been derived. In\cite{aminian2015capacity}, authors have considered a linear time invariant (LTI) Poisson channel for concentration based (CB) model, and then they derived some upper and lower bounds on the channel capacity. For this scenario, an optimum distribution for binary case has been found in \cite{6510564}. In all of these CB models, released concentration is chosen (according to a specific distribution) as the channel input, and the maximum mutual information between the released concentration and the number of received molecules is computed for the capacity evaluation.

The timing channels has been proposed in \cite{eckford2007nanoscale} as an alternative way of communications in nano-networks and its capacity has been studied. To this end, the first hitting time probability has been considered to detect the release time of molecules. Then, accordingly, lower bounds on the channel capacity have been derived. In addition, Monte Carlo methods have been employed for calculating the achievable rates.\cite{6620545} has studied timing channels with energy constraints. Timing molecular channels with drifts, which turns to an additive inverse Gaussian channel, has been introduced in \cite{6191345}, where capacity lower and upper bounds are presented for the memoryless case. In \cite{6949028}, the capacity of a channel with additive memoryless inverse Gaussian noise has been obtained in asymptotic regime. Finally in \cite{DBLP:journals/corr/FarsadMEG16}, assuming a limited life time for molecules, the authors have considered different detectors (first hitting time and average time) and have computed lower and upper bounds in each case (which have different noise characteristics).

Both CB and timing based modulations (TB), individually, have been proved to be useful in MC. The question here is that if it is beneficial to use joint TB and CB modulations. This question is challenging because their effects are correlated. Diffusion equation shows a strong correlation between effects of the concentration released and passage time of the molecules in the channel. In this work, our aim is to quantify these effects and describe corresponding channel properties. In fact, we want to answer the question that how using joint concentration and timing can improve the process of data transmission. In order to compute bounds on the channel capacity analytically, we assume that the transmitter chooses the release time among $m$ discrete times instead of choosing from a continuous range.  Our model uses different release times  and different levels of molecules in transmitter to capture their effects on the transmission rate. In the receiver, the observation time (symbol period) is divided into some sub-intervals and based on the number of molecules received in different sub-intervals, the decision is made on the time and the level of molecules released by the transmitter at the related symbol transmission. We assume a memoryless channel, where residue molecules from previous time slots are removed from the environment using for example a specific enzymes and imposing a guard interval between successive transmissions \cite{DBLP:journals/corr/FarsadMEG16}. We propose three lower bounds on the channel capacity of the JTAC modulation depending on how the counted number of molecules in different observation sub-intervals in the receiver is used as follows:\newline
1) For each sub-interval, we compute the mutual information between the number of molecules received in the sub-interval with the channel inputs (the release time and level of concentration). The maximum of these mutual information provides a lower bound on the capacity.
\newline 2) Using the received molecules in total observation sub-intervals, the level of concentration is determined. Then, given the released concentration, the mutual information between the number of molecules received in each sub-interval and the release time is computed. The maximum of these mutual information terms is used to derive the second lower bound on the channel capacity.
\newline 3) Again like second bound, using the total received molecules, the level of concentration is determined. Then, given the concentration, the mutual information between the difference of the number of molecules received in adjacent sub-intervals and the release time, for each adjacent sub-interval, is computed, and the maximum of these mutual information terms is used to derive the third lower bound.

For each of above lower bounds, we derive a closed form expression. We also compute an upper bound on the channel capacity of the JTAC modulation. To this end, for each observation sub-interval, we compute an upper bound on the mutual information between the number of received molecules and the channel input (release time and concentration level) using the symmetric KL divergence introduced in\cite{aminian2015capacity}. To see how tight the derived bounds are, we also numerically compute the capacity of JTAC channel using Blahut-Arimoto algorithm.

Our numerical results show that lower bounds on channel capacity are tight specifically for the environments with large diffusion coefficient. Also our results demonstrate that using JTAC modulation significantly increases achievable rates compared to conventional concentration based (CB) and timing based (TB) modulations as it discussed in numerical section. More specifically, for the number of observation sub-intervals 20 and the number of distinct release time of 10 and in an environment with diffusion noise parameter equal to 2, we achieve up to 1.5 bits (\%50) higher rates than the CB. Also we see that in environments with lower \l`evy noise parameter, $c$, timing based modulation has a larger part in total achievable rates in channel. Also From the example considered, JTAC provides more robust transmission compared to the conventional CB. That is, in our schemes capacity falls more slowly when environmental noise parameter increases compared to CB in higher values of $c$.

The rest of paper is organized as follows: in Section~\ref{sec:model}, we describe the system model. In Section~\ref{sec:lower}, we provide lower bounds on the capacity of the channel (achievable rates) presenting the three schemes stated above. Section~\ref{sec:upper} provides an upper bound on the channel capacity. In Section~\ref{sec:numerical}, numerical results are provided for the derived bounds. And finally in Section~\ref{sec:numerical}, some concluding remarks are presented.\newline
\textbf{Notation}: We denote random variables with upper case letters and their realizations with corresponding lower case letters. We use $f(.)$ to represent probability density function (PDF) of continuous and mixed Random Variables. the probability of an event A is donated by $Pr(A)$, and the probability mass function (PMF) of a discrete random variable $X$ is donated by $P_{X}(x)$. $\erf(x)$ is used to show error function given by $\frac{2}{\sqrt{\pi}} \int_{0}^{x} e^{-t^2} dt$. For simplicity, throughout paper, $h(.)$ represents differential entropy  of a continuous random variable or entropy of a discrete random variable based on its argument. $\log(.)$ is used to denote logarithm in natural basis. Basis for all other cases is specified in the related context. $D_{KL}$ denotes Kullback-Leibler divergence metric. And finally $\lambda_0$ is used to denote noise molecules in the environment.\footnote{This effect can be due to excessive molecules of the same type as the transmitted molecules that exist in diffusion environment from other sources. }
\section{System Model}\label{sec:model}
\begin{figure}[tb]
  \centering
  \includegraphics[width=9cm]{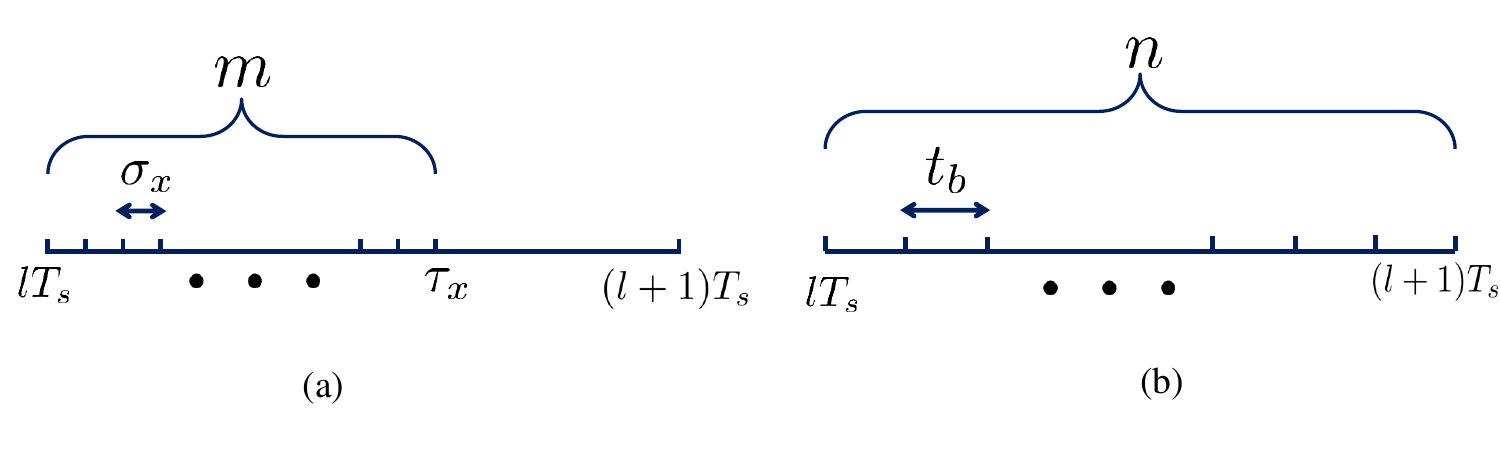}
  \caption{Time scheduling at the (a) Transmitter and (b) Receiver.}
  \label{fig:time}
\end{figure}

\begin{figure}[tb]
  \centering
  \includegraphics[width=7cm]{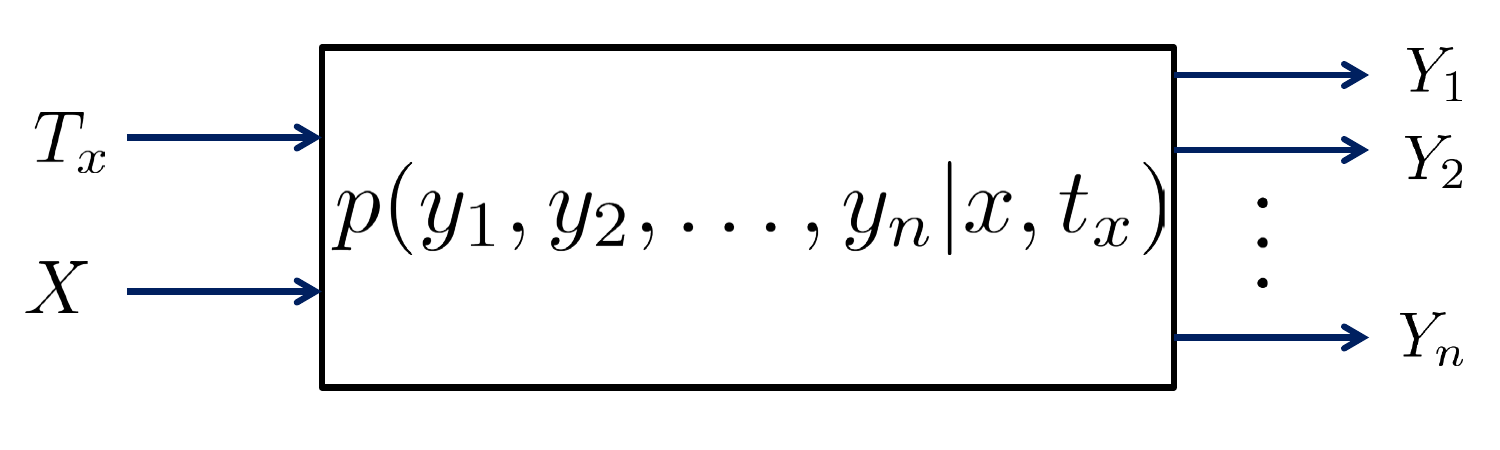}
  \caption{Channel model.}
  \label{fig:channel}
\end{figure}

Now, we describe the system model and its constraints. We consider a point to point MC system as follows:
\newline Transmitter: Transmitter is an exact concentration transmitter that completely controls the intensity and the time of released molecules.
We have both average and maximum  constraints on the released concentration (denoted by $X$),
\begin{equation}
E\{X\}\leq E_{m}
 ,\,\,\,\,\, 0\leq X\leq M.
\end{equation}
Transmission is time slotted with duration $T_s$. To avoid ISI, the release time is restricted to be in interval $[l T_s,l T_s+\tau_x]$ in $l$-th time slot, where $\tau_x\leq T_s$ is a design parameter based on the level of ISI\cite{DBLP:journals/corr/FarsadMEG16}. We denote the release time by $T_x $ which is a discrete random variable with $ m$ possible levels. In fact we have $T_x=lT_s+j\sigma_x, \,\,\,\,\,0\leq j\leq m-1$. The release time is assumed to have finite number of levels for the sake of tractability of capacity analysis. As $m$ increases, we get closer to the continuous timing channels, in which the release time can be any time in the transmission interval.
In each channel use, transmitter selects a time $T_x=lT_s+j\sigma_x$ and concentration $X=x$ consistent with (1). As said before, $T_x$ is selected in a time interval with duration $\tau_x$ which $\tau_x\leq T_s$. \cite{DBLP:journals/corr/FarsadMEG16} (see Fig. \ref{fig:time}). Also note that since our channel model is a memoryless channel and our analysis are independent of a specific time slot, in the following, without loss of generality, we assume $l=0$, i.e., $T_x$=$j\sigma_x$.
\newline Receiver: We consider an ideal receiver that absorbs any molecule that hits its surface. Each observation time slot at the receiver is divided in $n$ sub-intervals each with duration $t_b$. Receiver counts the number of molecules that hit it in every $t_b$ seconds. That is, we have $n$ observations in each time slot where $n=\frac{T_s}{t_b}$. The $i$-th observation (i.e., related to the number of molecules counted in the $i$-th sub-interval) is shown by $Y_i$ (see Fig. \ref{fig:channel}). We assume that the molecules not arriving at the receiver in the duration of corresponding time slot, disappear, which is possible for example by injecting some specific enzymes in the environment \cite{DBLP:journals/corr/FarsadMEG16}. Also note that the time taken by a molecule to hit the receiver is distributed according to a L\'evy distribution\cite{DBLP:journals/corr/FarsadYECG14}.
The probability that each released molecule falls in the specific sub-interval $i$ (1$\leq i\leq n $) with the assumption that the transmitter has released at time $j\sigma_x$ is computed by integrating PDF of L\'evy distribution in the corresponding sub-interval.\footnote{$p_{i j}$ is the solution of Fick's Diffusion equation.}
\begin{equation}\label{shw1}
p_{i j}=\int_{(i-1)t_b}^{it_b} L(j\sigma_x,c) dt,
\end{equation}
where
\begin{equation}\label{shw}
L(j\sigma_x,c)=\sqrt{\frac{c}{2\pi (t-j \sigma_x)^{3}}}\exp(\frac{-c}{2(t-j \sigma_x)}),\,\,t-j \sigma_x\geq 0.
\end{equation}
$c$ is L\'evy distribution parameter, which is defined as $\frac{d^2}{2D}$\label{lev} where $D$ is the diffusion coefficient of the environment and $d$ is the distance between transmitter and receiver. As $c$ increases (due to increase in $d$ or decrease in $D$) the environment becomes more noisy and transmission gets more unreliable.
From \eqref{shw1} and \eqref{shw}, we have:
\begin{equation}
    p_{i j}=\erf(\frac{\sqrt{c}}{\sqrt{2 ((i-1)t_b-j\sigma_x)}})-\erf(\frac{\sqrt{c}}{\sqrt{2 (it_b-j\sigma_x)}}).
\end{equation}
Based on the generalization of Bernoulli trials, if we define $A=Pr(Y_1=y_1,...,Y_n=y_n|T_x=t_x,X=x)$ as the probability that $y_i$ molecules fall in the $i$-th sub-interval $i=1,...,n$, given $T_x=t_x$ and $X=x $ we have $A=\frac {x !}{y_1!y_2!...y_n!(x-M_1)!}\prod_{i=1}^{n} p_{i j}^{y_i}(1-\sum_{i}^{}p_{i j})^{x-M_1}$,
where $M_1=y_1+...+y_n$. For large values of $x$, assuming that $\frac{x}{T_s}$ equals to a constant value, $\lambda$, we can use Poisson approximation as
\begin{equation}\label{finalf}
A=e^{-\lambda_{1 j}}\frac{(\lambda_{1 j})^{y_1}}{y_1!}\times e^{-\lambda_{2 j}}\frac{(\lambda_{2 j})^{y_2}}{y_2!}...e^{-\lambda_{n j}}\frac{(\lambda_{n j})^{y_{n}}}{y_{n}!},
\end{equation}
where we have $ \lambda_{i j}=x p_{i j}, 0\leq j \leq m-1$, and $p_{i j}$ is defined in \eqref{shw1}.
So, the channel transition probability will be
\begin{align*}
&Pr(Y_1=y_1,Y_2=y_2,...,Y_n=y_n|X=x,T_x=t_x)\\
&=\prod_{i}^{} Pr(Y_i=y_i|X=x, T_x=t_x),
\end{align*}
where $t_x=j\sigma_x$ ,\, $0 \leq j\leq m-1$. Then from \eqref{finalf} we have:
\begin{equation}\label{eq1-0}
Pr(Y_i=y_i|X=x,T_x=t_x)=e^{(-x p_{i j})}\frac{(x p_{i j})^{y_i}}{y_i!}.
\end{equation}
\begin{remark} We neglect the impact of noise molecules in the environment throughout the paper (in contrast to L\'evy noise which is inherent to diffusion environment) except in part B of section \ref{sec:lower} and section \ref{sec:upper}. When considering the environmental noise molecules, the channel transition probability is the same as \eqref{finalf}, except that we have $\lambda_{i j}= M_1 p_{i j}+\lambda_0$.
	\end{remark}
\section{Lower bounds on Channel Capacity}\label{sec:lower}
In this section, we provide three lower bounds on the JTAC channel capacity.
\subsection{First lower bound}
 As stated before, in the first scheme, the receiver uses only the observation of a single sub-interval which has maximum mutual information with the input. This lower bound simply uses the information of one sub-interval and throws away information of other sub-intervals for the sake of the simplicity.
 \begin{theorem}
A lower bound on the JTAC channel capacity based on maximum of mutual information between input and the number of received molecules in each different sub-intervals is
 \begin{equation}\label{here-2}
C\geq \max_{i} I(X,T_x;Y_i)\geq \max_{i} R_i,
\end{equation}
where
\begin{flalign*}
 R_i&=\log m +\frac{1}{2}\log \frac{M}{\mu}+\log \sqrt{\pi}\erf{\sqrt{\mu}}+\alpha \mu \\
&-\log E_{m} p_i^{*} -m -\log k-\frac{1}{2}\log{2\pi e}-\frac{1}{2} \log p_i^{*}\\
&-\frac{1}{2}\bigg(\frac{4\sqrt{\frac{\mu}{12 M p_i^{*}}}\taninv (\sqrt{12 M p_i^{*}})+2\sqrt{\mu}\log(1+\frac{1}{12 p_i^{*} M})}{\sqrt{\pi}\erf{\sqrt{\mu}}}\bigg),
\end{flalign*}
where\,$\mu$\,is the solution of $\alpha=\frac{E_{m}}{M}=\frac{1}{2\mu}-\frac{e^{-\mu}}{\sqrt{\mu}\sqrt{\pi} {{\erf{\sqrt{\mu}}}}}$,
and
\begin{equation}\label{whatt}
p_{i}^{*}=\max_{j}p_{i j}.
    \end{equation}
    \end{theorem}
\begin{proof}
To prove \eqref{here-2}, the receiver determines the $i$-th sub-interval that maximizes $I(X,T_x;Y_i)$ for $ 1\leq i\leq n $. To this end, we derive the bounds on this mutual information using Lemmas 1 and 2 given below.
One strategy for computing lower bound on $I(X,T_x;Y_i)$ is to lower bound $h(Y_i)$ and upper bound $h(Y_i|X,T_x)$ in terms of input distributions and channel parameters.
For computing a lower bound on $h(Y_i)$, we borrow a technique from \cite{4729780}, and then we upper bound $h(Y_i|X,T_x) $ based on the results on the entropy of Poisson random variable. Final result of the lower bound is in terms of $h(X,T_x)$, $ E\{X\}$ and $p_{i j}$, which can be maximized under constraints (1) and (2). This concludes explicit lower bounds on $I(X,T_x;Y_i)$.
To compute a lower bound on $h(Y_i)$ we need a variant of data processing theorem from \cite{korner1}.
\begin{lemma}\cite[Lemma 3.11]{korner1}
For any distributions $P$ and $Q$ on $X$ and any stochastic matrix $w={w(y|x):x\in X,y\in Y }$ we have $ D_{KL}(Pw||Qw)\leq D_{KL}(P||Q)$, i.e., every processing on distributions $P$ and $Q$ decreases their $KL $ divergence.
\end{lemma}
Now, using above lemma we find a lower bound on $h(Y_i)$ in terms of input distribution of channel which could be maximized by choosing appropriate joint distribution on $(X,T_x)$ which we see in further discussions.
 \begin{lemma}
A lower bound on entropy of output of JTAC Channel is as follows:
 \begin{align}
         h(Y_i)\geq h(X,T_x)-\log(\eta p_{i}^{*})-m-\log(k).\label{lemm2}
         \end{align}
    \end{lemma}
\begin{proof}
 We use Lemma 1 by choosing an arbitrary input distribution $Q(.)$ and a specific input distribution with pdf
  \begin{equation}\label{rare}
      P=f(x,T_x=j\sigma_x)=\frac{k}{\eta p_{i j}}e^{-\frac{1}{\eta}x},
\end{equation}
where $k$ is selected such that $f(X, T_x=j\sigma_x)$ sums to one. In fact, we have $k=(\sum_{j=1}^{m} \frac{1}{p_{i j}})^{-1}$. From \eqref{rare}, we have
\begin{align}
D_{KL}&(Q(T_x,X)||P(T_x,X))\notag\\
&=\int_{0}^{\infty}\sum_{i=0}^{m} Q(T_x,X)\log(\frac{Q(T_x,X)}{\frac{k}{\eta p_{i j}}e^{-\frac{x}{\eta }}})dX\label{rare0}\\
&=-h(Q(X,T_x)+1+\sum_{j=1}^{m} Q(T_x=j\sigma_x) \log(\eta p_{i j}).\notag
\end{align}
Now let the the PMF of the output of Poisson channel with input distributions $Q$ and $P$ be shown by $ P_{Q}^{o}(.)$ and $P_{G}^{o}(.)$, respectively.
From \eqref{rare}, we have: 
\begin{equation}
    P_{G}^{o}=P_{Y_i}(y_i)=\sum_{t_x}^{}\int_{0}^{\infty}\frac{k}{\eta p_{i j}}\ e^{\frac{-1}{\eta}x}p(y_i|x,t_x)dx,
\end{equation}
Using \eqref{eq1-0}, it can be shown that:
\begin{equation}
P_{G}^{o}=\sum_{j=1}^{m}\frac{k}{p_{i j}}(\frac{1}{\eta p_{i j}+1})(\frac{\eta p_{i j}}{\eta p_{i j}+1})^{y}.
\end{equation}
Thus using the definition of $KL$ divergence for $P_{Q}^{o}$ and $P^{o}_{G}$ :
\small
\begin{align}
&D_{KL}(P_{Q}^{o}(.)||P_{G}^{o}(.))=\sum_{y=0}^{\infty} P_{Q}^{o}(y_i)\log(\frac{P_{Q}^{o}(y_i)}{\sum_{i=1}^{m} \frac{k}{\eta p_{i j}}\frac{1}{\eta p_i+1}(\frac{\eta p_i}{\eta p_i+1})^{y}})\nonumber\\
&\!\!\!\!\!\!\!\!=-h(Y_i)-\sum_{y=0}^{\infty} P_{Q}^{o}(y_i)\log({\sum_{j=1}^{m} \frac{k}{\eta p_{i j}}\frac{1}{\eta p_{i j}+1}(\frac{\eta p_{i j}}{\eta p_{i j}+1})^{y}}).\label{rare1}
\end{align}
\normalsize
Applying Lemma 1 to \eqref{rare0} and \eqref{rare1} results in:
\begin{align}
h(Y_i)\geq& h(X,T_x)-\sum_{j=1}^{m} Q(T_x=j\sigma_x)\log(\eta p_{i j})-1 \label{hg0}\\
&-\sum_{y=0}^{\infty} P_{Q}^{o}(y_i)\log({\sum_{j=1}^{m} \frac{k}{\eta p_{i j}}\frac{1}{\eta p_{i j}+1}(\frac{\eta p_{i j}}{\eta p_{i j}+1})^{y}}).\notag
\end{align}
Now, we upper bound two sums in the above equation to conclude a simple form for the lower bound on $h(Y_i).$ By using $ \log(x)\leq \,\,(x-1)$, we obtain
\begin{flalign}
     &\sum_{y=0}^{\infty} P_{Q}^{o}(y_i)\log({\sum_{j=0}^{m-1} \frac{1}{\eta p_{i j}}\frac{1}{\eta p_{i j}+1}(\frac{\eta p_{i j}}{\eta p_{i
    j}+1})^{y}}) \label{rare2} \\
     &\leq \sum_{y=0}^{\infty} P_{Q}^{o}(y_i)\Big(\sum_{j=0}^{m-1} \frac{1}{\eta p_{i j}}\frac{1}{\eta p_{i j}+1}(\frac{\eta p_{i j}}{\eta p_{i j}+1})^{y} -1\Big)\\&=\sum_{j=0}^{m-1}\sum_{y=0}^{\infty}P_{Q}^{o}(y_i).\frac{1}{\eta p_{i j}}\frac{1}{\eta p_{i j}+1}(\frac{\eta p_{i j}}{\eta p_{i j}+1})^{y}-1 \notag\\
     &\overset{(a)}{\leq} \sum_{j=0}^{m-1}\frac{1}{\eta p_{i j}}\sum_{y=0}^{\infty}y.\frac{1}{\eta p_{i j}+1}(\frac{\eta p_{i j}}{\eta p_{i j}+1})^{y}-1\notag\\
     &\overset{(b)}{=}\sum_{j=0}^{m-1}\frac{1}{\eta p_{i j}}.\eta p_{i j}-1=m-1,\notag
\end{flalign}
where (a) follows from the fact that $Y_i$ is an integer valued random variable and (b) is due to using the average of Geometric distribution.
\newline
Next, we upper bound $\sum_{j=1}^{m}Q(t_x=j\sigma_x)\log(\eta p_{i j})$ in \eqref{hg0}. We know that $\log(p_{i j})$ is a monotonic function and above sum has a form of average of a random variable with probability distribution $ Q(t_x)$. This sum can be upper bounded as:
\begin{equation}\label{rare3}
 \sum_{j=1}^{m}Q(t_x=j\sigma_x)\log(\eta p_{i j})\leq \log(\eta p_{i}^{*}),
 \end{equation}
where $p_{i}^{*}$ is defined in \eqref{whatt} in Combining \eqref{rare2} and \eqref{rare3} results in \eqref{lemm2}.
\end{proof}
\begin{lemma}
An upper bound on the conditional entropy of the number of received molecules in a specific sub-interval $i$ is
\begin{align}
    h(Y_i|X,T_x)\leq & \frac{1}{2} \log(2\pi e )+\frac{1}{2}E\{\log(p_{i}^{*} X)\}+\nonumber\\
     &\frac{1}{2}E\{\log(1+\frac{1}{12p_{i}^{*} X})\}.\label{here-1}
\end{align}
\end{lemma}
\begin{proof}
We have:
\begin{equation}\label{here-3}
    h(Y_i|X,T_x)=\sum_{j=0}^{m} h(Y_i|X,T_x=j\sigma_x)P_{T_x}(j\sigma_x),
\end{equation}
where each entropy term can be upper bounded as follows (using \cite[Theorem 8.6.5]{10.5555/1146355}):
since summing an independent uniform random variable with an arbitrarily random variable \,\,\,increases its variance, and by using the entropy of a Gaussian random variable, we have the following upper bound
\begin{flalign*}
    &h(Y_i|X,T_x=j\sigma_x)\leq =\frac{1}{2}\log(2 \pi e)+\frac{1}{2}\log(E\{p_{i j} X+\frac{1}{12}\}).
\end{flalign*}
Using the Jensen's inequality, we have:
\begin{align}
&h(Y_i|X,T_x=j\sigma_x)\leq\frac{1}{2}\log(2 \pi e)\nonumber \\&+\frac{1}{2}E\{\log(p_{i j} X)\}+\frac{1}{2}E\{\log(1+\frac{1}{12 p_{i j} X})\}.\label{magical1}
 \end{align}
Now, consider $P_{T_x}^{*}$ as the distribution that maximizes \eqref{here-3}. So,
\begin{align*}
  &h(Y_i|X,T_x)\leq \sum_{j=0}^{m}\big(\frac{1}{2} \log(2\pi e )+\frac{1}{2}E\{\log(p_{i j} X)\}\\
  &+\frac{1}{2}E\{\log(1+\frac{1}{12p_{i j} X})\}\big)
 P_{T_x}^{*}(j\sigma_x).
 \end{align*}
Noting that the right side of above equation is an increasing function of $p_{i j}$, we achieve \eqref{here-1}.
\end{proof}
Using \eqref{lemm2} and \eqref{here-1} and $I(X,T_x;Y_i)=h(Y_i)-h(Y_i|X,T_x)$, it can be seen:
\begin{align}
&I(Y_i;X,T_x)\geq h(X,T_x)-\log(\eta p_{i}^{*})-m-\log(k)\label{here13}\\
&-(\frac{1}{2} \log(2\pi e )+\frac{1}{2}E\{\log(p_{i}^{*} X)\}+\frac{1}{2}E\{\log(1+\frac{1}{12p_{i}^{*}
X})\}).\nonumber
 \end{align}
Since the above lower bound is valid for any joint distribution of $T_x$ and $X$, we should maximize this expression over $f(X,T_x)$ to conclude an explicit lower bound on $I(Y_i;T_x,X)$.
A suboptimal solution is to make an independence assumption on $ X $ and $T_x$ and maximize each entropy to maximize the lower bound. In other words, since an exponential distribution and a uniform distribution maximize entropy for a mean constrained and finite alphabet random variables, respectively, we can set $f(X,T_x=j\sigma_x)=\frac{\gamma}{m} e^{-\frac{-x}{\eta}}$ to maximize \eqref{here13}. But with this assumption, the term $E\{\log(X)\}$ cannot be derived explicitly. To overcome this issue, we maximize $h(X)-\frac{1}{2} E\{\log(X)\}$ under constraints in (1). This is a standard convex optimization problem that can be solved using Lagrange multipliers.\newline
The objective function in this case is
 \begin{align*}
     J(f)=&-\int f(x) \log f(x) -\frac{1}{2}\int \log(x) f(x)+\Gamma_0 \int f(x)\\
          &+\Gamma_1 \int x f(x),
\end{align*}
 where $\Gamma_0$ and $\Gamma_1$ are Lagrange Multipliers. We have
 \begin{equation*}
     \frac{\partial J(f)}{\partial f(x)}=-\log(f(x)-1-\frac{1}{2} \log(x)+\Gamma_0+\Gamma_1 x=0,
 \end{equation*}
 solving for $f(x)$, we have $f(x)=\frac{k_1}{\sqrt{x}} e^{k_2 x}$.
Applying constraints (1), we conclude
\begin{equation}\label{herefg}
     f(x)=\frac{\sqrt{\mu}}{\sqrt{M \pi x}\erf{\sqrt{\mu}}}.
\end{equation}
Using this distribution, we compute an upper bound on\allowdisplaybreaks \quad $E\{\log(1+\frac{1}{12p^{*}_i X})\}$ using \cite[eq.(43)]{4729780}.
\small
 \begin{align*}
   E\{ \log(1+\frac{1}{12p^{*}_i X} ) \} &=\int_{0}^{M}
    \log(1+\frac{\frac{1}{12p^{*}_i}}{x}).\frac{\sqrt{\mu}}{\sqrt{M \pi x}\erf{\sqrt{\mu}}}e^{\frac{-\mu x}{A}}\notag \\
    &\leq \int_{0}^{M} \log(1+\frac{\frac{1}{12p^{*}_i}}{x}).\frac{\sqrt{\mu}}{\sqrt{M \pi x}\erf{\sqrt{\mu}}}dx
    \end{align*}
    \normalsize
    \begin{align}
&=\bigg(\frac{4\sqrt{\frac{\mu}{12 M p_i^{*}}}\ \taninv (\sqrt{12 M p_i^{*}})+2\sqrt{\mu}\log(1+\frac{1}{12 p_i^{*} M})}{\sqrt{\pi}\erf{\sqrt{\mu}}}\bigg).\label{herefg1}
\end{align}
By substituting \eqref{herefg} and \eqref{herefg1} in \eqref{here13}, we obtain
\begin{align*}
&I(X,T_x;Y_i)\geq \max_{i} \{\log m +\frac{1}{2}\log \frac{M}{\mu}+\log \sqrt{\pi}\erf{\sqrt{\mu}}+\alpha \mu\\
&-\log E_{m} p_i^{*} -m-\log k-\frac{1}{2}\log{2\pi e}-\frac{1}{2} \log p_i^{*}\\
&-\frac{1}{2}\bigg(\frac{4\sqrt{\frac{\mu}{12 M p_i^{*}}}\ \taninv (\sqrt{12 M p_i^{*}})+2\sqrt{\mu}\log(1+\frac{1}{12 p_i^{*} M})}{\sqrt{\pi}\erf{\sqrt{\mu}}}\bigg) \}.
\end{align*}
where $\alpha$ and $\mu$ are defined in \eqref{here-2}.
This concludes the lower bound on the channel capacity given in \eqref{here-2}.
\end{proof}
\subsection{Second lower bound}
In this scheme, the receiver uses sum of received molecules in all sub-intervals to detect the transmitted concentration, while for the release time detection, the sub-interval with maximum mutual information with the released time is used.
\begin{theorem}
Using the sum of molecules counted in different sub-intervals, the JTAC channel capacity is lower bounded as
\begin{equation}\label{prf1}
    C\geq \max (R_1,R_2),
\end{equation}
where
\small
\begin{align}
&R_1= \log_2(m)-\log(c^{'})-E_{m} \phi -\log(E_{m} p^{*})-\log(m)-1\nonumber\\
&-\log(2\pi e)-\log(k^{'})+ \log(12)+\max_{i} h(\textrm{Poisson}(M \tilde{p_{i}})),\label{eq01-1}\\
&R_2=\log_2(m)-\log(c^{'})-E_{m} \phi -\log(E_{m} p^{*})-m\label{eq01-2}\nonumber\\
&-\log(2\pi e)-\log(k^{'})+ \log(12)+\max_{i} h(\textrm{Poisson}(M \tilde{p_{i}})),
 \end{align}
 \normalsize
and $\phi$ is computed by solving \eqref{tkhere} given in Appendix A. Also note that $c^{'}$ is computed using $\phi$ derived in \eqref{tkhere} and \eqref{eq4-3},
 $Ei(.)$ is a special function called exponential integral function\cite{olver2010nist} defined in \eqref{here3003} which is used to evaluate $\phi$, and
\begin{align}
k^{'}&=\frac{1}{\sum_{j} \frac{1}{p_{j}^{'}}},\label{here29}\\
p_{j}^{'}&=\sum_{i=0}^{m}p_{i j},\label{eq3-5}\\
p^{*}&=\max_{j} p_{j}^{'}.\label{eq3-1}\\
\tilde{p_{i}}&=\min_{j}p_{i j}
\end{align}
\end{theorem}
\begin{proof}
The proof is provided in Appendix A.
\end{proof}
\begin{remark}
It will be shown in Section \ref{sec:numerical} that this scheme outperforms the one in Theorem 1 in some cases. In fact for large $m$, we have $R_2$ as tighter lower bound, while for small $m$, $R_1$ achieves higher rates.
\end{remark}
\subsection{Third lower bound}
The third scheme uses the difference of received molecules in adjacent sub-intervals to detect the release time, while the detection of concentration is similar to the second scheme (i.e., using the sum of received molecules). That is, the receiver counts the increase or decrease of received molecules in adjacent sub-intervals in order to detect transmitted time $T_x$ conditioned on knowing the transmitted concentration level. Since this scheme could be of practical interest because of its simplicity, this bound could also give some insight on achievable rates by practical receivers.\newline
The distribution of the difference of two independent Poisson random variables contains modified Bessel function \cite{papoulis2001probability}, and as a result computing the entropy becomes intractable in this case. Thus, we consider Gaussian approximation for Multinomial distribution as follows:
\begin{align}
    Pr(Y_1=y_1,...,Y_n=y_n|X=x,T_x=t_x)\label{here44}\\
    =\frac{x!}{y_1! y_2!...y_n!}p_{1j} p_{2j}...p_{nj}=\prod_{i} Pr(y_i|x,t_x),\nonumber
\end{align}
where $t_x=j\sigma_x$ and
\begin{equation}\label{here114}
    Pr(Y_i=y_i|X=x,T_x=t_x)= \frac{e^{\frac{-1}{2xp_{ij}}( y_i-xp_{ij})^{2}}}{\sqrt{2\pi x p_{ij}}}.
\end{equation}
 Now, we have
 \begin{equation}\label{here2020}
     I(X,T_x;Y_1,...,Y_n)=I(X;Y_1,...,Y_n)+I(T_x;Y_1,...,Y_n|X).
 \end{equation}
      To detect the concentration (the first term in \eqref{here2020}), we use the lower bound in \eqref{eq3-3}, which is obtained by using received molecules in all observation sub-intervals. For the release time (the second term in \eqref{here2020}), based on the Markov chain,
 \begin{equation*}
        (X, T_x)\rightarrow (Y_{1},...,Y_{n})\rightarrow f(Y_{1},...,Y_{n})=Y_{i}-Y_{i-1},
 \end{equation*}
 we have
 \begin{equation*}
     I(T_x;Y_1,...,Y_n|X)\geq I(T_x;Y_i-Y_{i+1}|X).
 \end{equation*}
So we can find the following lower bound (i.e., achievable rate) on the JTAC channel capacity:
\begin{equation*}
    C\geq I(X;Y_1,...,Y_n)+\max_{i}I(T_x;Y_i-Y_{i-1}|X),
\end{equation*}
which is computed in the following theorem.
\begin{theorem}
The JTAC channel capacity using the difference of received molecules in adjacent sub-intervals can be lower bounded as:
\begin{flalign}
\!\!\!\!C& \geq \max_{i}\frac{1}{2 m}(\log(M)\erf{\frac{M}{\sqrt{u}}})\nonumber \\
&-\frac{1}{2 m \sqrt{\pi u}}M \Hypergeometric{2}{2}{\frac{1}{2},\frac{1}{2}}{\frac{3}{2},\frac{3}{2}}{\frac{-M^2}{u}}\nonumber\\
&+\sum_{k=0}^{r} \sum_{j}^{}a_{ij}z_k m_{k i j}+\frac{1}{2 m} \sum_{j}^{}(\frac{1}{2}\log(2 \pi e  q^{'}_{ij})+\frac{1}{2}(2 \pi u e)\nonumber\\
& -\log(\eta p^{*})-m-\log(k^{'})-\frac{1}{2m} \sum_{j}^{}\log(p_{j}^{'}).\label{diff}
\end{flalign}
where $\Hypergeometric{2}{2}{\frac{1}{2},\frac{1}{2}}{\frac{3}{2},\frac{3}{2}}{x}$ is the Generalized Hypergeometric function \cite{olver2010nist}, $m_{k i j}$ and $a_{i j}$ will be clarified in \eqref{here1000} and \eqref{here1001}, and $u$ is specified according to the distribution used in \eqref{here1002}.
\end{theorem}
\begin{proof}
 A proof is given in Appendix B.
 \end{proof}
In next section we consider problem of finding Upper bounds on channel capacity.
\section{Upper bounds On Channel Capacity}\label{sec:upper}
To derive an upper bound on the channel capacity, first we obtain an upper bound on mutual information between inputs and number of received molecules in a specific sub-interval.
\begin{lemma}
For $i$-th sub-interval at the receiver we have
\begin{equation}
    I(X,T_x;Y_i)\leq \frac{E_{m} p_{i}^{*}}{M}(M-E_{m}) \log(\frac{M p_{i}^{*}}{\lambda_0} +1),
\end{equation}
where $Y_i$ is the number of molecules received in $i$-th sub-interval and $p^{*}_{i}$ is defined in \eqref{whatt}.
\end{lemma}
\begin{figure*}
    \begin{subfigure}{.5\textwidth}
    \centering
    \includegraphics[width=7.5cm]{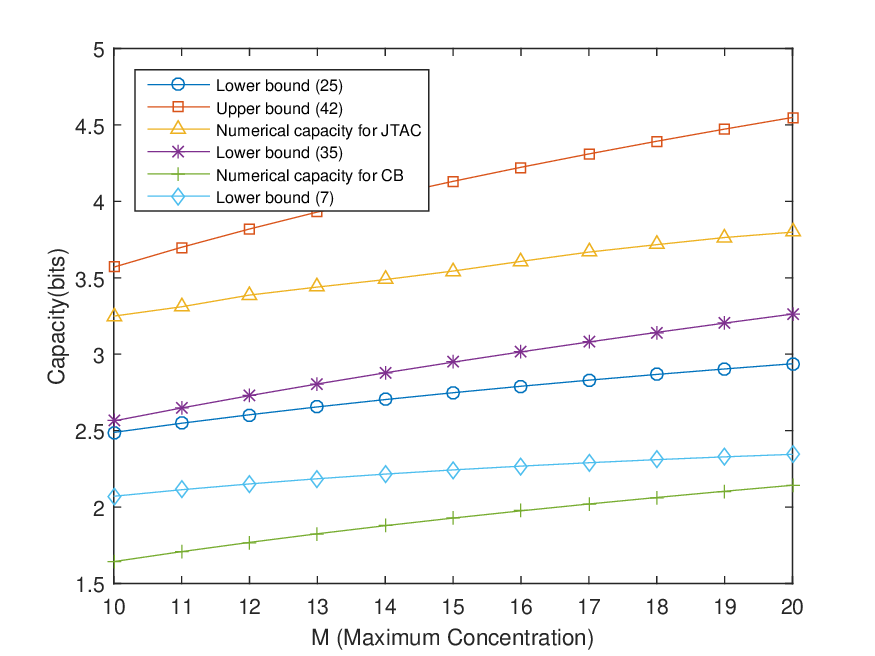}
    \caption{$c=2$}
    \label{figure1}
    \end{subfigure}
    \begin{subfigure}{.5\textwidth}
     \centering
    \includegraphics[width=7.5cm]{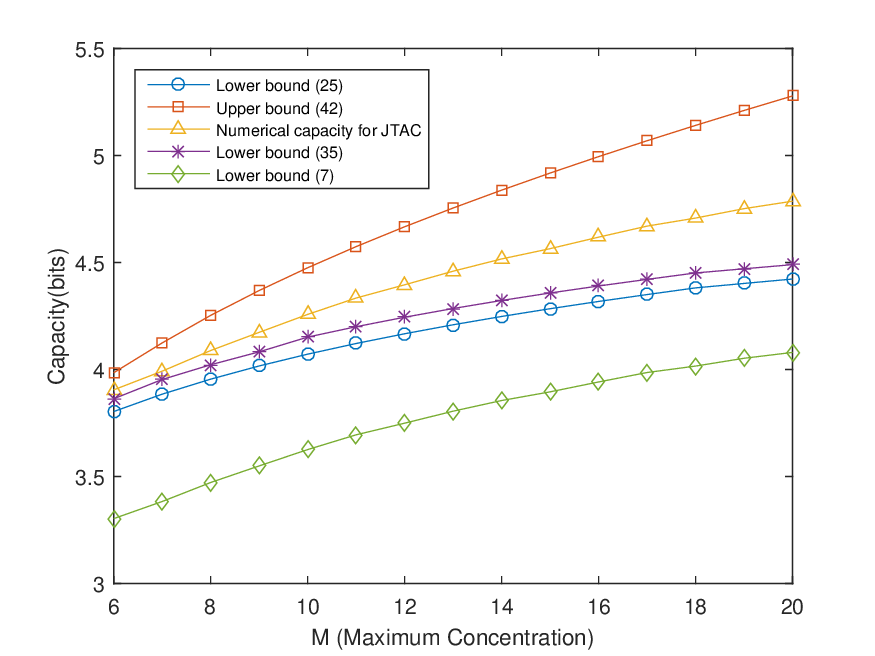}
    \caption{ $c=.1$ }
   \label{figure6}
   \end{subfigure}
   \caption{Lower bounds \eqref{here-2}, \eqref{prf1}, and \eqref{diff}, upper bound and numerical capacity (Blahut-Arimoto) of JTAC and CB channels for $m=10$, $n=20$ and $\xi=\frac{E_m}{M}=\frac{1}{5}$. }
   \label{fig11}
\end{figure*}
\begin{proof}
We use the symmetric KL divergence, which has been defined in \cite{aminian2015capacity} as:
\begin{equation}
 D_{sym}(P||Q)=D_{KL}(P||Q)+D_{KL}(Q||P).
\end{equation}
Now, we set
\begin{equation*}
 P=p(x,t_x,y_i),\quad Q=p(x,t_x)p(y_i),
\end{equation*}
and obtain an upper bound on $I(X,T_x;Y_i)$. It is easy to show 
\begin{align}
    D_{sym} (P||Q)=&\mathop{\mathbb{E}_{p(x,t_x,y_i)}\log(p(y_i|x,t_x))}\notag \\
    &-\mathop{\mathbb{E}_{p(x,t_x)p(y_i)} \log(p(y_i|x,t_x))}.\label{hipp}
\end{align}
Using the fact that $E\{ E\{Y|X,T_x\}\}=E\{Y|X\}$, and from \eqref{eq1-0} we obtain 
$$\mathop{\mathbb{E}_{p(x,t_x,y_i)}}\log p(y_i|X,T_x=t_x)=E\{(p_{i j} X+\lambda_0 )\log(p_{i j} X+\lambda_0)\},$$ $$\mathop{\mathbb{E}_{p(x,t_x)p(y_i)}}\log p(y_i|X,T_x=t_x)=E\{p_{i j} X+\lambda_0 \}E\{\log(p_{i j} X+\lambda_0)\}.$$
Combining above expressions with \eqref{hipp} results in,
\begin{align*}
   D_{sym^c} =&E\{(p_{i j} X+\lambda_0 )\log(p_{i j} X+\lambda_0)\mid T_x=t_x\} \notag\\
       &-E\{p_{i j} X+\lambda_0 \}E\{\log(p_{i j} X+\lambda_0)\mid T_x=t_x\}\label{func1},
   \end{align*}
   using $E\{ E\{Y|X,T_x\}\}=E\{Y|X\}$ we have:
   \begin{align}
   D_{sym}&=\sum_{j=1}^{m} Pr(T_x=t_x)E\{(p_{i j} X+\lambda_0 )\log(p_{i j} X+\lambda_0)\}\notag\\
   &-E\{p_{i j} X+\lambda_0 \}E\{\log(p_{i j} X+\lambda_0)\}.
   \end{align}
Then, noting $I(X,T_x;Y_i)\leq D_{sym}$, total upper bound is:
\begin{align*}
   I(X,T_x,Y_i)\leq& \sum_{j=1}^{m} Pr(T_x=t_x)E\{(p_{i j} X+\lambda_0 )\log(p_{i j} X+\lambda_0)\}\\
   &-E\{p_{i j} X+\lambda_0 \}E\{\log(p_{i j} X+\lambda_0)\}.
\end{align*}
Since a binary random variable maximizes expression in \eqref{func1}, using similar steps as in \cite{aminian2015capacity} for $E_{m} \leq \frac{M}{2}$, we get
\begin{equation}\label{herewhe}
    \max(D_{sym})=\frac{E_{m} p_{i j}}{M}(M-E_{m}) \log(\frac{M p_{i j}}{\lambda_0} +1).
\end{equation}
So, we have:
\small
\begin{align*}
    I(X,T_x;Y_i)&\leq \sum_{j=1}^{m} Pr(T_x=t_x)\frac{E_{m} p_{i j}}{M}(M-E_{m}) \log(\frac{M p_{i j}}{\lambda_0} +1)
    \end{align*}
    \normalsize
    \begin{flalign}
    &\overset{a}{\leq} \frac{E_{m} p_{i}^{*}}{M}(M-E_{m}) \log(\frac{M p_{i}^{*}}{\lambda_0} +1),\label{finalle}
\end{flalign}
where (a) follows from \eqref{whatt} and the fact that \eqref{herewhe} is an increasing function of $p_{i j}$.
\end{proof}
\begin{theorem}
 The capacity of the JTAC channel is upper bounded as:
 \begin{equation}\label{fghm}
         C\leq \sum_{i}^{} \frac{E_{m} p_{i}^{*}}{M}(M-E_{m}) \log(\frac{M p_{i}^{*}}{\lambda_0} +1).
\end{equation}
\end{theorem}
\begin{proof}
Using memoryless property of channel, we have:
 \begin{flalign*}
     I(X,T_x;Y_1,...,Y_n)&\leq \sum_{i}^{}I(X,T_x;Y_i).
 \end{flalign*}
Substituting \eqref{finalle} in the above equation we obtain \eqref{fghm}.
 \end{proof}
\section{Numerical Results}\label{sec:numerical}
\begin{figure}
    \centering
    \includegraphics[width=7.5cm]{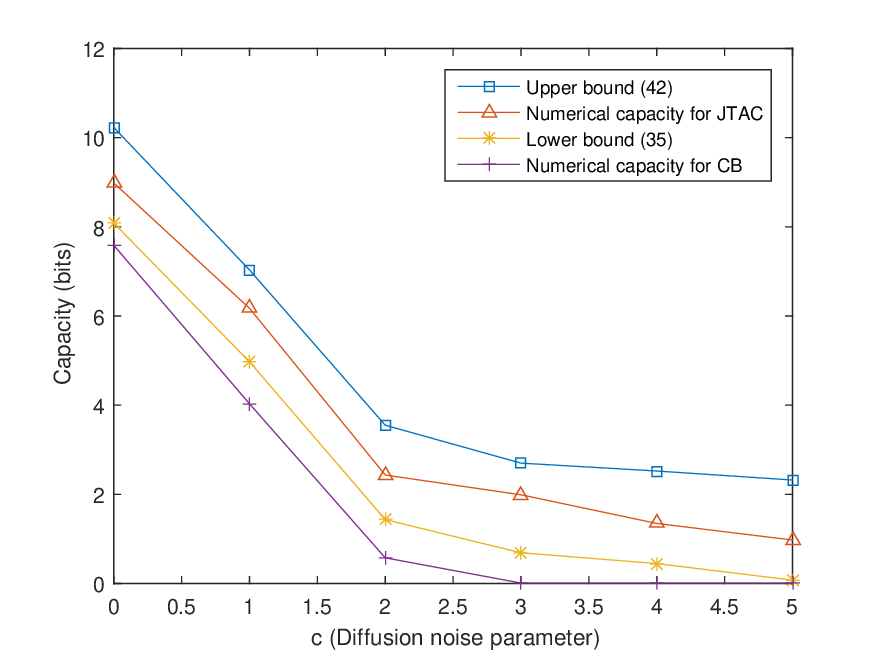}
    \caption{Effect of $c$ (L\'evy noise parameter) on transmission rates with $M=10$, $m=5$, $n=10$.}
    \label{figure2}
\end{figure}

\begin{figure}
    \centering
    \includegraphics[width=7.5cm]{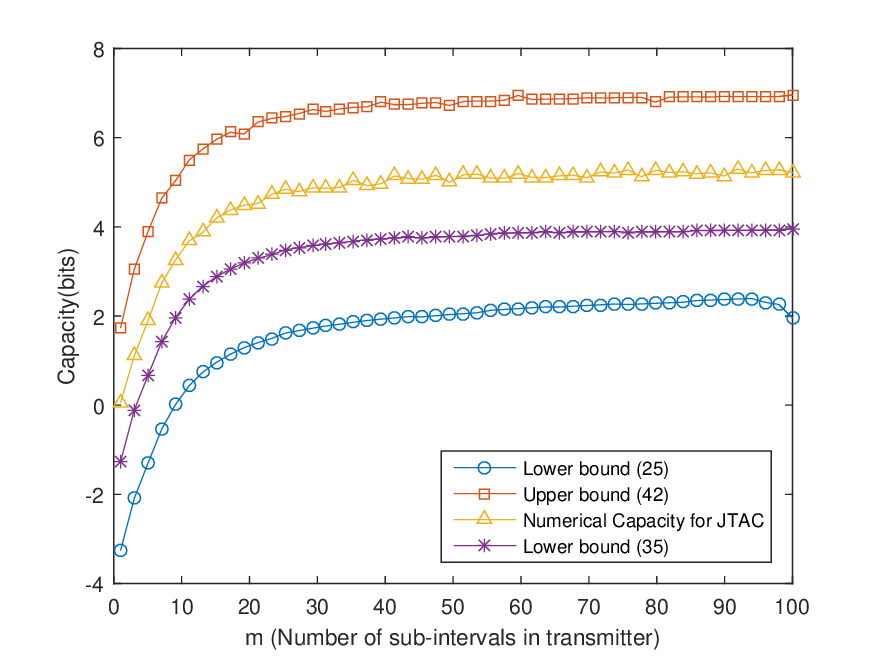}
    \caption{Effect of $m$ (number of sub-intervals in transmitter) on transmission rates with $c=1$, $M=15$, $n=10$.}
    \label{figure4}
\end{figure}

\begin{figure}
    \centering
    \includegraphics[width=7.4cm]{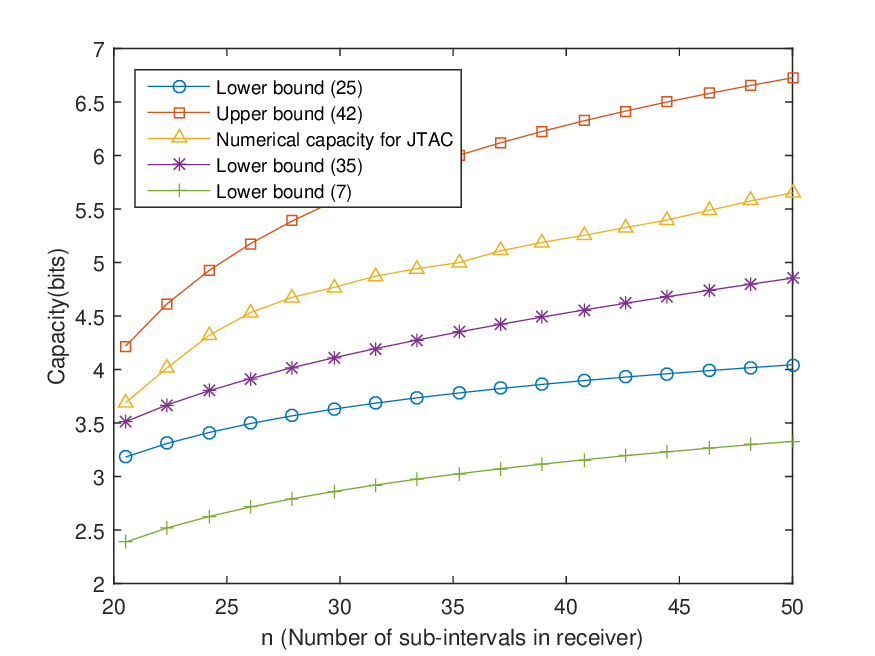}
    \caption{Effect of $n$ (number of sub-intervals in receiver) on transmission rates with $c=1$, $M=15$, $m=10$.}
    \label{figure5}
\end{figure}

\begin{figure*}
    \begin{subfigure}{.33\textwidth}
    \centering
    \includegraphics[width=6.7cm]{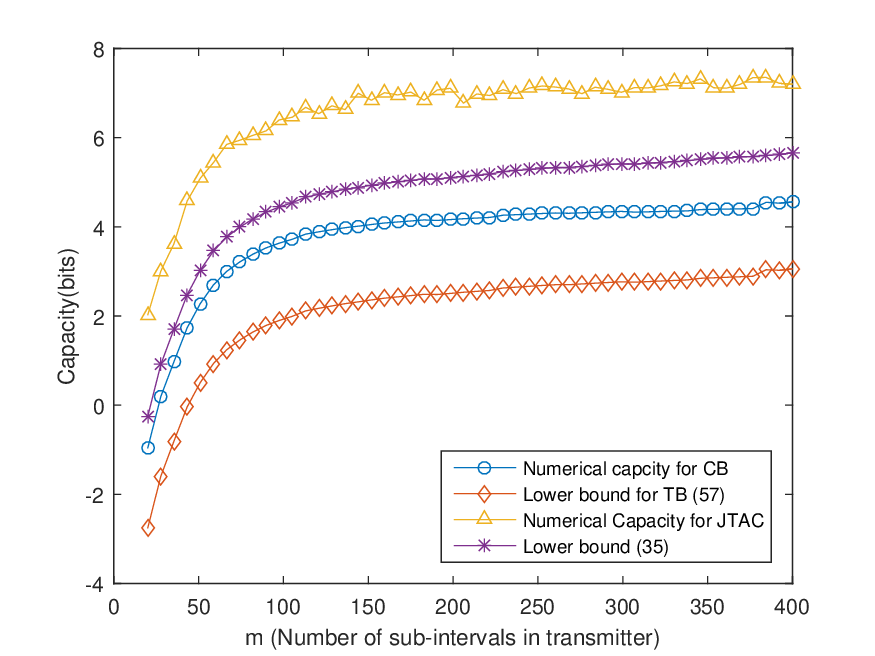}
    \caption{$c=1$, $M=15$, $n$=40}
    \label{figfin1}
    \end{subfigure}
    \begin{subfigure}{.33\textwidth}
     \centering
    \includegraphics[width=6.7cm]{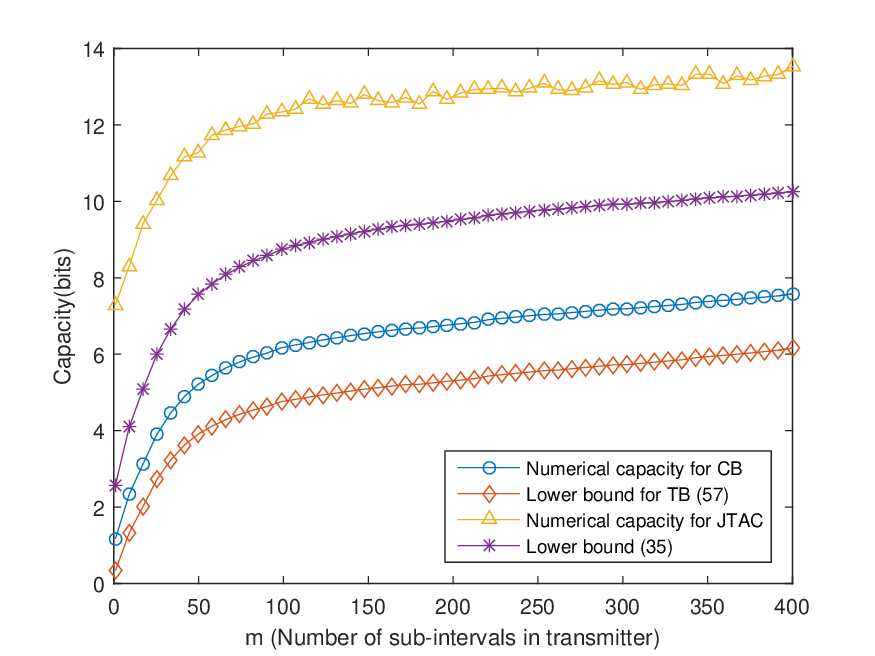}
    \caption{ $c=0.1$, $M$=15, $n$=40 }
   \label{figfin2}
   \end{subfigure}
   \begin{subfigure}{.33\textwidth}
   \centering
   \includegraphics[width=6.7cm]{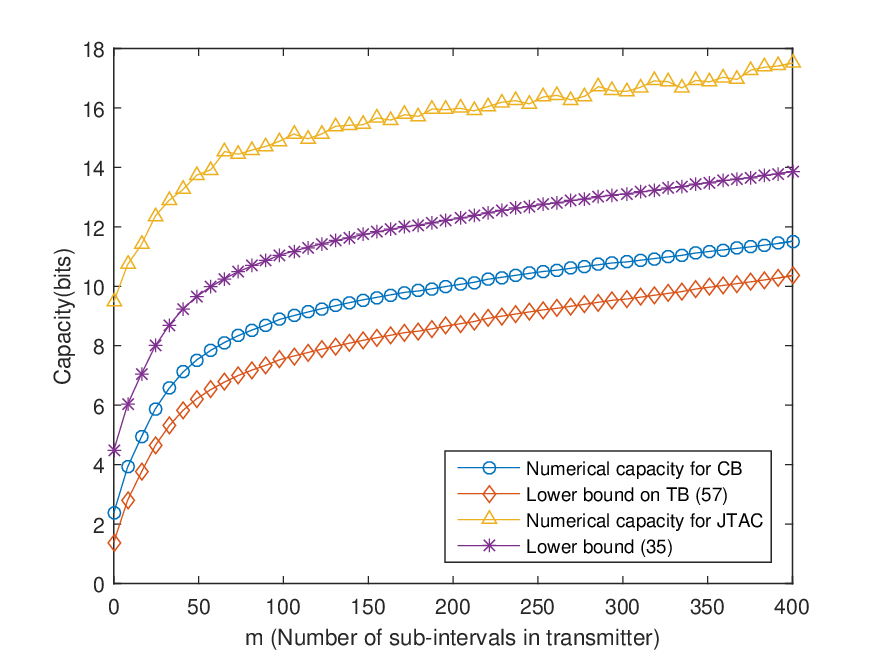}
   \caption{$c=0.1$, $M=25$, $n=80$}
   \label{figfin3}
   \end{subfigure}
   \caption{Lower bounds \eqref{here-2}, \eqref{prf1}, and \eqref{diff}, upper bound and numerical capacity (Blahut-Arimoto) of JTAC and CB channels for $\xi=\frac{E_m}{M}=\frac{1}{5}$. }
   \label{fig}
\end{figure*}

\begin{table}[t]
   \centering
\begin{tabular}{| l|c|c |c |c |c |c |c |}
    \hline
    $c$ & 0.1 &  1& 2& 3& 4&5\\
   \hline
   $D (\mu m ^{2}/s)$&
    4800 & 480& 240& 160& 120 & 96\\
    \hline
\end{tabular}
\caption{Setup for numerical simulations for some typical values of Diffusion coefficient from \cite{10.1093/nar/gkp889} and $d (\mu m)=21.91$.}
\label{table:1}
\end{table}
In this section, we provide numerical evaluations of the proposed lower and upper bounds on the capacity of JTAC channel. We also numerically compute JTAC capacity using Blahut-Arimoto algorithm to examine how tight the bounds are. Moreover, we discuss the improvements compared with the concentration based (CB) modulation and timing based (TB) modulation. For CB, the information is only coded in $X$, keeping $T_x$ constant. For TB, the transmitter releases a fixed concentration of molecules in a sub-interval chosen according to the input message. Again by denoting the number of molecules received in $i-th$ sub-interval by $Y_i$, we consider $\max_{i}I(T_x;Y_i|X=x)$, as a lower bound on the capacity of TB.\footnote{Since $I(T_x;Y_1,...,Y_n|X=x)$ is of the form of a side information we could not use directly use Blahut-Arimoto to compute TB capacity numerically\cite{1365218}.} For CB channel, we use Blahut-Arimoto algorithm to compute its capacity numerically.\newline Fig. \ref{figure1} depicts the proposed lower bounds and the upper bound versus the maximum concentration ($M$) and compares them with the numerically evaluated capacity of the JTAC and the CB channels. It is evident that the capacity and the bounds increase with $M$, as expected. Also it is seen that the lower bound which is based on the difference of received molecules in adjacent sub-intervals (third scheme, given in \eqref{diff}) provides higher achievable rates than the others. Since realizing receivers that detect the concentration of molecules in a limited number of time-slots seems to be more feasible compared to the receivers that have to detect the arrival time of molecules continuously, this suggests that third scheme could be considered for designing practical receivers to detect the release time. An important conclusion from Fig. \ref{figure1} is that there is a relative large gap between the capacities of CB channel and JTAC. This indicates that using both time and concentration, to encode the message, results in significant improvement on achievable rates up to \%50 (1.5 bits) compared to using only concentration. Fig. \ref{figure6} depicts all bounds versus $M$ with a lower diffusion noise parameter, i.e., $c$. From Fig. \ref{figure1} and Fig. \ref{figure6} it can be seen that the proposed bounds become tighter at lower $c$. Also it can be observed that in lower $c$, the capacity increases more rapidly with $M$ and obviously achieves higher values than in environments with larger $c$. To study the effect of $c$ on our bounds and the channel capacity, we consider a setup according to Table \ref{table:1}, where different values of diffusion coefficient are considered for a fixed distance between the transmitter and the receiver to simulate environments with different L\'evy diffusion noise parameters, i.e., $c$. Using this setup, Fig. \ref{figure2} considers the effect of $c$ on the channel capacity. We expect that with increasing $c$, the capacity decreases due to the decrease in the diffusion coefficient of the environment (see Section \ref{sec:model} and Table \ref{table:1}), which is the case for all achievable rates and the upper bound depicted in Fig. \ref{figure2}. Also Fig. \ref{figure2} shows that with increasing $c$, the capacity of CB channel for values of $c\geq 2$ falls more rapidly compared with the capacity (and even achievable rates) of JTAC channel. This suggests that using JTAC modulation provides a more robust strategy compared to conventional CB modulations in larger values of $c$ (in this case 2) and JTAC is less sensitive to diffusion noise.\newline Fig. \ref{figure4} depicts the achievable rates and the  upper bounds on the channel capacity versus the number of sub-intervals at the transmitter ($m$). As we increase $m$, we get closer to the continuous timing channel and thus we could achieve higher rates. But an important observation from Fig. \ref{figure4} is that increasing the number of sub-intervals beyond a point which for this parameter setup is 50, has a little impact on the channel capacity. We observe similar results in other simulations with different environmental noise parameters; however, the point of saturation increases in less noisy channel (smaller value of $c$). This further motivates us considering the more practical discrete time-slotted timing channel (compared with the continuous timing channel). Fig. \ref{figure5} shows the effect of the number of observation times at the receiver on the capacity. It is seen that by increasing the number of observation times in receiver, we can achieve higher rates as we expected. Finally Fig. \ref{figfin1} to \ref{figfin3} compares timing based rates (TB) with other bounds and channel capacity versus $m$. It is seen by comparing Fig. \ref{figfin2} and \ref{figfin3} that increase in $n$ increases timing rates (thanks to the more accurate detection in receiver). And Fig. \ref{figfin1} and \ref{figfin2} compares TB rates with other bounds in environments with different $c$. As it seen from Fig. \ref{figfin2}, the gap between lower bound on TB and CB has been reduced compare to Fig. \ref{figfin1}, which indicates that in environments with small $c$, timing rate has larger portion in total rate of the channel and CB is more resistant to noise $c$ compared to TB.
\section{conclusion}
In this paper, we introduced JTAC modulation for molecular diffusion channel, in which the information is modulated in both the release time and concentration of transmitted molecules. In order to analyze the JTAC performance in comparison with prior modulation schemes, more specifically concentration based (CB) modulation, we considered its capacity and derived three lower bounds and one upper bound on the capacity. We numerically evaluated the capacity of JTAC and CB modulation using Blahut-Arimoto algorithm and obtained a lower bound on the capacity of TB modulation. Our results indicate that the lower bound for the JTAC channel based on detecting the difference of the numbers of molecules in adjacent sub-intervals at the receiver provides tighter lower bound compared to the two other lower bounds. It is also observed that the capacity and achievable rates of the JTAC channel increase with the increase of the number of discrete release times at the transmitter or the number of observation sub-intervals at the receiver, up to a saturation point. The value of the saturation point increases, with increase in the number of sub-interval at the transmitter or by decrease in diffusion noise  parameter ($c$). Our results further indicate that the JTAC modulation significantly improves the achievable rates compared to CB and TB modulations. For example, as shown in Fig. \ref{figure1} we could achieve up to 1.5 bits per symbol higher rates compared to CB. And finally our numerical results indicated that in higher values of $c$ JTAC falls less rapidly with increasing $c$, which suggest that using JTAC could provide more robust strategy for transmission in the environment with large $c$ compared to CB. In our analysis, we neglected the ISI assuming a time gap between successive symbol intervals and left the analysis in the presence of ISI for future work. Another interesting problem is considering effect of flow on our results and its joint effect on both CB and TB. Also, Considering more realistic conditions, for example non-ideality of the transmitter and examining its effects are another area of future work.

\bibliographystyle{IEEEtran}
\bibliography{main1}

\begin{thebibliography}{10}
\providecommand{\url}[1]{#1}
\csname url@samestyle\endcsname
\providecommand{\newblock}{\relax}
\providecommand{\bibinfo}[2]{#2}
\providecommand{\BIBentrySTDinterwordspacing}{\spaceskip=0pt\relax}
\providecommand{\BIBentryALTinterwordstretchfactor}{4}
\providecommand{\BIBentryALTinterwordspacing}{\spaceskip=\fontdimen2\font plus
\BIBentryALTinterwordstretchfactor\fontdimen3\font minus
  \fontdimen4\font\relax}
\providecommand{\BIBforeignlanguage}[2]{{%
\expandafter\ifx\csname l@#1\endcsname\relax
\typeout{** WARNING: IEEEtran.bst: No hyphenation pattern has been}%
\typeout{** loaded for the language `#1'. Using the pattern for}%
\typeout{** the default language instead.}%
\else
\language=\csname l@#1\endcsname
\fi
#2}}
\providecommand{\BIBdecl}{\relax}
\BIBdecl

\bibitem{DBLP:journals/corr/FarsadYECG14}
N.~{Farsad}, H.~B. {Yilmaz}, A.~{Eckford}, C.~{Chae}, and W.~{Guo}, ``A
  comprehensive survey of recent advancements in molecular communication,''
  \emph{IEEE Comm. Surveys Tutorials}, vol.~18, no.~3, pp. 1887--1919, 2016.

\bibitem{DBLP:journals/corr/FarsadMEG16}
N.~{Farsad}, Y.~{Murin}, A.~W. {Eckford}, and A.~{Goldsmith}, ``Capacity limits
  of diffusion-based molecular timing channels with finite particle lifetime,''
  \emph{IEEE Trans. on Molecular, Biological and Multi-Scale Comm}, vol.~4,
  no.~2, pp. 88--106, June 2018.

\bibitem{gohari2016information}
A.~Gohari, M.~Mirmohseni, and M.~Nasiri-Kenari, ``Information theory of
  molecular communication: Directions and challenges,'' \emph{IEEE Trans. on
  Molecular, Biological and Multi-Scale Comm}, vol.~2, no.~2, 2016.

\bibitem{einolghozati2011capacity}
A.~Einolghozati, M.~Sardari, A.~Beirami, and F.~Fekri, ``Capacity of discrete
  molecular diffusion channels,'' in \emph{Proc. IEEE Intern. Symp. on Inf.
  Theory}, 2011, pp. 723--727.

\bibitem{6648629}
A.~{Einolghozati}, M.~{Sardari}, and F.~{Fekri}, ``Design and analysis of
  wireless communication systems using diffusion-based molecular communication
  among bacteria,'' \emph{IEEE Trans. on Wireless Comm}, vol.~12, no.~12, pp.
  6096--6105, December 2013.

\bibitem{aminian2015capacity}
G.~Aminian, H.~Arjmandi, A.~Gohari, M.~Nasiri-Kenari, and U.~Mitra, ``Capacity
  of diffusion-based molecular communication networks over {LTI}-{Poisson}
  channels,'' \emph{IEEE Trans on Molecular, Biological and Multi-Scale Comm},
  vol.~1, no.~2, pp. 188--201, 2015.

\bibitem{6510564}
B.~{Atakan}, ``Optimal transmission probability in binary molecular
  communication,'' \emph{IEEE Comm. Letters}, vol.~17, no.~6, June 2013.

\bibitem{eckford2007nanoscale}
A.~W. Eckford, ``Nanoscale communication with brownian motion,'' \emph{41st
  Annual Conf. on Inf. Sciences and Systems}, pp. 160--165, 2007.

\bibitem{6620545}
C.~{Rose} and I.~S. {Mian}, ``Signaling with identical tokens: Lower bounds
  with energy constraints,'' in \emph{2013 IEEE Intern. Symp. on Inf. Theory},
  July 2013, pp. 1839--1843.

\bibitem{6191345}
K.~V. {Srinivas}, A.~W. {Eckford}, and R.~S. {Adve}, ``Molecular communication
  in fluid media: The additive inverse {Gaussian} noise channel,'' \emph{IEEE
  Trans. on Inf. Theory}, vol.~58, no.~7, pp. 4678--4692, July 2012.

\bibitem{6949028}
H.~{Li}, S.~M. {Moser}, and D.~{Guo}, ``Capacity of the memoryless additive
  inverse {Gaussian} noise channel,'' \emph{IEEE Journal on Selected Areas in
  Comm}, vol.~32, no.~12, pp. 2315--2329, Dec 2014.

\bibitem{4729780}
A.~{Lapidoth} and S.~M. {Moser}, ``On the capacity of the discrete-time
  {Poisson} channel,'' \emph{IEEE Trans. on Inf. Theory}, vol.~55, no.~1, pp.
  303--322, Jan 2009.

\bibitem{korner1}
I.~Csiszár and J.~Körner, \emph{Information Theory: Coding Theorems for
  Discrete Memoryless Systems}.\hskip 1em plus 0.5em minus 0.4em\relax
  Cambridge University Press, 2011.

\bibitem{10.5555/1146355}
T.~M. Cover and J.~A. Thomas, \emph{Elements of Information Theory}.\hskip 1em
  plus 0.5em minus 0.4em\relax USA: Wiley-Interscience, 2006.

\bibitem{olver2010nist}
F.~W. Olver, D.~W. Lozier, R.~F. Boisvert, and C.~W. Clark, \emph{NIST handbook
  of mathematical functions}.\hskip 1em plus 0.5em minus 0.4em\relax Cambridge
  university press, 2010.

\bibitem{papoulis2001probability}
A.~Papoulis and U.~Pillai, ``Probability, random variables and stochastic
  processes,'' \emph{McGraw-Hill}, 2001.

\bibitem{10.1093/nar/gkp889}
\BIBentryALTinterwordspacing
R.~Milo, P.~Jorgensen, U.~Moran, G.~Weber, and M.~Springer, ``Bionumbers—the
  database of key numbers in molecular and cell biology,'' \emph{Nucleic Acids
  Research}, vol.~38, no. suppl-1, pp. D750--D753, 10 2009. [Online].
  Available: \url{https://doi.org/10.1093/nar/gkp889}
\BIBentrySTDinterwordspacing

\bibitem{1365218}
F.~{Dupuis}, W.~{Yu}, and F.~M.~J. {Willems}, ``Blahut-arimoto algorithms for
  computing channel capacity and rate-distortion with side information,'' in
  \emph{IEEE Intern. Symp. on Inf. Theory (ISIT)}, 2004.

\bibitem{DBLP:journals/corr/abs-cs-0607075}
\BIBentryALTinterwordspacing
C.~Nair, B.~Prabhakar, and D.~Shah, ``On entropy for mixtures of discrete and
  continuous variables,'' \emph{CoRR}, vol. abs/cs/0607075, 2006. [Online].
  Available: \url{http://arxiv.org/abs/cs/0607075}
\BIBentrySTDinterwordspacing

\bibitem{8437888}
M.~{Cheraghchi}, ``Expressions for the entropy of binomial-type
  distributions,'' in \emph{IEEE Intern. Symp. on Inf. Theory (ISIT)}, June
  2018.

\bibitem{huber2008entropy}
M.~F. Huber, T.~Bailey, H.~Durrant-Whyte, and U.~D. Hanebeck, ``On entropy
  approximation for {Gaussian} mixture random vectors,'' \emph{IEEE Intern.
  Conf. on Multisensor Fusion and Integration for Intelligent Systems}, pp.
  181--188, 2008.

\end{thebibliography}
\appendices
\section{Proof of \eqref{prf1}}
Having access to the observations $Y_1,Y_2,...,Y_n$, the receiver uses their sum, to decode the transmitted concentration $X$ and then conditioned on the transmitted $X$ the receiver uses the sub-interval which its observations has the maximum mutual information with the transmitted time $T_x$. Thus, we have:
\begin{align}
I(X,T_x&;Y_1,...,Y_n)=I(X;Y_1,...,Y_n)+I(T_x;Y_1,...,Y_n|X),\nonumber\\
   &\overset{(a)}{\leq} I(X;Y_1+...+Y_n)+\max_{i} I(T_x;Y_i\mid X).\label{heretk20}
\end{align}
where (a) follows since $(X,T_x)\rightarrow(Y_1,...,Y_n)\rightarrow f(Y_1,...,Y_n)=Y_1+...+Y_n$ forms a Markov chain and mutual information is non negative. We present proof in three steps.\newline
\text{Step 1) Lower bound on $I(X;Y_1+Y_2+,...+Y_n)$}:
 Using the fact that conditioned on $T_x$ and $X$ the receiver observations $Y_i$ are independent Poisson random variable, we conclude that:
\begin{equation}\label{mahgh}
\!p(y=y_1+...+y_n|X=x,T_x=j \sigma_x)=e^{(-x p_{j}^{'})}\frac{(x p_{j}^{'})^{y}}{y!},
\end{equation}
Now consider:
\begin{equation}\label{here1}
\!\!I(X;Y_1+...+Y_n)=h(Y_1+...+Y_n)-h(Y_1+...+Y_n|X),
\end{equation}
We lower bound the first term and upper bound the second term in \eqref{here1} to conclude a lower bound on $I(X;Y_1+...+Y_n)$.
By using similar steps as taken in proof of Lemma 2 we have:
\begin{equation}
    h(Y_1+...+Y_n)\geq h(X,T_x)-\log(\eta p^{*})-m-\log(k^{'})\label{eq3-0}.
\end{equation}
For computing the upper bound on second term in \eqref{here1} note that from \eqref{mahgh} we have:
\begin{equation*}
    Pr(Y=y_1+...+y_n|X=x)=\sum_{j}^{}e^{(-x p_{j}^{'})}\frac{(x p_{j}^{'})^{y}}{y!}Pr(T_x=j\sigma_x),
\end{equation*}
To compute the second term in \eqref{here1}, we should compute the entropy of above distribution.
To do this, we define an auxiliary random variable $\theta$ as
\[ \theta=
    \begin{cases}
    $$L_1 \sim \textrm{Poisson}(x\,p_{1}^{'})$$ &\text{with probability } $$Pr(T_x=\sigma_x)$$\\
    \vdots\\
   $$ L_m \sim \textrm{Poisson}(x\,p_{m}^{'})$$ &\text{with probability }$$Pr(T_x=m \sigma_x)$$
    \end{cases}
    \]
Using the results from entropy of mixtures \cite{DBLP:journals/corr/abs-cs-0607075} we have:
\begin{equation}\label{here3}
    h(Y_1+Y_2+...+Y_n|X)=h(T_x)+\sum_{j}^{}h(L_j)Pr(T_x=j\sigma_x).
\end{equation}
Based on upper bound on Poisson distribution in \eqref{here-1}, we upper bound \eqref{here3} to conclude:
\begin{align}
\!\!h(Y_1+&...+Y_n|X)\leq h(T_x)+\sum_{j}^{}Pr(T_x=j\sigma_x)\Big(\frac{1}{2} \log(2\pi e )\nonumber\\
&+\frac{1}{2}E\{\log((p_{ j}^{'}X)\})+\frac{1}{2}E\{\log(1+\frac{1}{12p_{j}^{'} X}\}\Big).\label{eq3-2}
\end{align}
Substituting \eqref{eq3-0} and \eqref{eq3-2} in \eqref{here1}, we have:
\begin{flalign}
\!\!\!\!\!\!\!\!\!\!\!\!\!\!\!I(Y_1&+...+Y_n;X)\!\geq h(X,T_x)-h(T_x)-\log(\eta p^{*})-m \label{eq3-3}\\
&-\log(k^{'})-\sum_{j}^{}Pr(T_x=j\sigma_x)\Big(\frac{1}{2} \log(2\pi e )+\frac{1}{2}E\{\log((p_{j}^{'}X)\}\Big).\notag
\end{flalign}
Step 2) Lower Bound on $I(T_x;Y_1,...,Y_n|X)$:
It is easy to show that:
\begin{equation*}
   \max_{i} I(T_x,Y_i|X)\leq I(T_x;Y_1,...,Y_n|X),
\end{equation*}
So, we turn to compute (lower) bounds on $ \max_{i} I(T_x,Y_i|X)$
We have:
\begin{equation}\label{eq3-4}
    I(T_x,Y_i|X)=h(Y_i|X)-h(Y_i|X,T_X).
\end{equation}
In \eqref{here-1} we computed an upper bound on the second term of \eqref{eq3-4}. So, it suffices to lower bound $ h(Y_i|X)$ to obtain  a lower bound on $I(T_x,Y_i|X)$.
Using \eqref{eq1-0}
distribution of $Y$ conditioned on $X$ is computed as:
\begin{equation}\label{nbvc}
  Pr(Y_i=y_i|X=x)=\sum_{j}^{}Pr(T_x=j\sigma_x) e^{(-x p_{i j})}\frac{(x p_{i j})^{y_i}}{y_i!},
  \end{equation}
 To lower bound entropy of \eqref{nbvc} consider following mixture:
 \[ \theta^{'}=
    \begin{cases}
    $$z_1 \sim \textrm{Poisson}(x\,p_{1}^{'})$$ &\;\; \text{with probability}\,\, $$Pr(T_x=\sigma_x)$$\\
    \vdots\\
    $$z_j=v_j $$ & \;\; \text{with probability} \,\, $$Pr(T_x=j \sigma_x)$$\\
    \vdots\\
   $$ z_m =v_m $$ & \;\; \text{with probability}\,\,$$Pr(T_x=m \sigma_x)$$
    \end{cases}
    \]
    Since alphabets of this mixture are disjoint (in fact except its first component others are fixed and not random)
\begin{flalign}\label{here9720}
h(Y_i|X=x)=h(T_x)+Pr(T_x=j\sigma_x)h(L_{i j}),
\end{flalign}
where
\begin{flalign}
&L_{i j}\sim \text{Poisson}(p_{i j}x).\notag
\end{flalign}
Since $h(L_{i j})$ is an increasing function of its argument and using the fact that $p_{i j}\leq \tilde{p_{i}}$ we have:\allowdisplaybreaks
\begin{equation*}
    h(Y_i|X)\geq h(T_x)+h(\text{Poisson}(M \tilde{p_{i}})).
\end{equation*}
So total lower bound on $I(T_x;Y_i|X)$ is
\begin{flalign}
&I(T_x;Y_i|X)\geq h(T_x)+h(\text{Poisson}(M \tilde{p_{i}})) \label{eq4-6}\\
&-\sum_{j}^{}Pr(T_x=j\sigma_x)\big(\frac{1}{2} \log(2\pi e )+\frac{1}{2}E\{\log(p_{i j} X)\} \notag \\
&+\frac{1}{2}E\{\log(1+\frac{1}{12p_{i j} X})\}\big).
\end{flalign}
\textbf{Remark 3}: Using \eqref{here-1} and \eqref{here9720}, $I(T_x,Y_i|X=x)$ can be lower bounded.
From \eqref{here-1} for a specific event \{$X=x$\}, we have,
\begin{align*}
  h(Y_i|X=x,T_x)=\sum_{j}^{} Pr(T_x=j\sigma_x)(\frac{1}{2} \log(p_{i j}x+\frac{1}{12}),
\end{align*}
so using \eqref{here9720} and the fact that $\log$ is a monotonic increasing function we conclude
\begin{flalign*}
    &I(T_x;Y_i|X=x)\geq h(T_x)+\sum{}^{} Pr(T_x=j\sigma_x)h(L_{i j})\nonumber\\&-\sum_{j}^{} Pr(T_x=j\sigma_x)(\frac{1}{2} \log(p_{i j}x+\frac{1}{12})).
\end{flalign*}
It is easy to see that this expression can be lower bounded as:
\begin{equation}\label{figtime}
    I(T_x;Y_i|X=x)\geq \log_{2}(m)+\frac{1}{m}\sum_{}^{}h(L_{i j})-\log(p_{i}^{*}x+\frac{1}{12}).
\end{equation}
This lower bound gives specifies how using timing in JTAC give us additional information (rates) compared to detecting only concentration\footnote{In fact additional rates that is achieved in addition to $I(X,Y_1,..,Y_n)$ }.\newline
\text{Step 3) Deriving lower bounds on the channel capacity}:
By using bounds in \eqref{heretk20}, \eqref{eq3-3}, and \eqref{eq4-6}, we have the following lower bound on mutual information of inputs and outputs:
\begin{flalign}
&I(T_x,X;Y_1,...,Y_n)\geq h(X,T_x)+\max_{i}h(\text{Poisson}(M \tilde{p_{i}})) \notag \\
&-\log(\eta p^{*})-m-\log(k^{'}) \notag \\
&-\sum_{j}^{}Pr(T_x=j\sigma_x)\big(\frac{1}{2} \log(2\pi e )+\frac{1}{2}E\{\log(p_{i j} X)\}\notag \\
&+\frac{1}{2}E\{\log(1+\frac{1}{12p_{i j} X})\}\big)-\sum_{j}^{}Pr(T_x=j\sigma_x)\notag \\
&\big(\frac{1}{2} \log(2\pi e)
+\frac{1}{2}E\{\log((p_{j}^{'}X)\})+\frac{1}{2}E\{\log(1+\frac{1}{12p_{ j}^{'} X}\}\big),\notag
\end{flalign}
which simplifies to\allowdisplaybreaks
\begin{flalign}
&I(T_x,X;Y_1,...,Y_n)\geq h(X,T_x)-\log(\eta p^{*})-m\\
&-\log(k^{'})-\log(2 \pi e)\notag
-\sum_{j}^{}Pr(T_x=j\sigma_x)\\ &\big(\frac{1}{2}\log(p_j^{'})-\frac{1}{2}\log(12p_{j}^{'})
+\frac{1}{2}E\{\log(12p_{j}^{'}X+1)\}\big)\notag \\
&-\max_{i}\sum_{j}^{} p(T_x=j \sigma_x)\big(\frac{1}{2}\log(p_{i j})-\frac{1}{2}\log(12p_{i j})\notag \\
&+\frac{1}{2}E\{\log(12p_{i j}X+1)\}\big)+h(\text{Poisson}(M \tilde{p_{i}})).\notag
\end{flalign}
Using $p^{*}\geq\max_{i}p_{i}^{*}$, we could lower bound the above expression to conclude the following simple form\footnote{Here we use the fact that the expectation of a random variable is less than or equal to its maximum realization.}:
\begin{align}
        I(T_x,X;Y_1,...&,Y_n)\geq h(X,T_x)-E\{\log(12p^{*}X+1)\}\nonumber\\
 &-\log(\eta p^{*})-m-\log(k^{'})-\log(2 \pi e)\nonumber\\
    & +\log(12)+\max_{i}h(\text{Poisson}(M \tilde{p_{i}})),\label{eq4-1}
 \end{align}
 It is easy to see that if we consider the noise molecules in \eqref{eq4-1} we obtain $I(T_x,X;Y_1,...,Y_n)\geq h(X,T_x)-E\{\log(12(p^{*}X+\lambda_{0})+1\}-\log(\eta p^{*})-m-\log(k^{'})-\log(2 \pi e)+ \log(12)+\max_{j}h(\text{Poisson}(M \tilde{p_{i}}))$.

By choosing appropriate distribution for $(X,T_x)$, we can maximize this lower bound.
Noting that $h(X,T_x)\leq h(X)+h(T_x)$, we use a uniform distribution for $T_x$ which maximizes its entropy in a finite range, and choose $X$ independent of $T_x$ and according to the distribution that maximizes  $h(X)-E\{\log (12p^{*}X+1)\}$.
Using Lagrange multipliers, $X$ has the following form \cite{10.5555/1146355}:
\begin{equation}\label{eq4-2}
    f_X(x)=\frac{c^{'}}{bx+1}e^{\phi x}, \quad\quad b=12p^{*}.
\end{equation}
where $\phi$ is computed such that constraint (1) is satisfied and $c^{'}$ is such that this distribution integrates to one. Thus, we have:
\begin{equation}\label{tkkk}
    \frac{e^{\phi M}-1}{bk}+\frac{e^{\frac{-\phi }{b}}(Ei(\frac{\phi}{b})-Ei(\phi M+\frac{\phi }{b}))}{b^{2}}=\frac{E_{m}}{c^{'}},
\end{equation}
\begin{equation}\label{eq4-3}
    \frac{e^{\frac{-k}{b}}}{b}\big(Ei(\phi M+\frac{\phi}{b})-Ei(\frac{\phi}{b})\big)=\frac{1}{c^{'}},
\end{equation}
where $Ei(.)$ is special function\footnote{Exponential integral function} and is defined as:
\begin{equation}\label{here3003}
  Ei(x)=-\int_{-x}^{\infty}\frac{e^{-t}}{t}dt.
\end{equation}
 Using \eqref{tkkk} and \eqref{eq4-3}, $\phi$ is computed by solving the following equation:
 \begin{equation}\label{tkhere}
    \frac{e^{\frac{-\phi}{b}}}{b}\big(Ei(\phi M+\frac{\phi}{b})-Ei(\frac{\phi}{b})\big)=\frac{e^{\phi M}-1}{b\phi(E_{m}-\frac{1}{b})},
 \end{equation}
Also it is straightforward to see that:
\begin{equation}\label{eq4-4}
 h(X)=-\log(c^{'})+E\{\log(12p^{*}X+1)\}-\eta \phi.
\end{equation}
Substituting \eqref{eq4-2}, \eqref{eq4-3}, and \eqref{eq4-4} in \eqref{eq4-1} results in:
\begin{align*}
     I(&T_x,X;Y_1,...,Y_n)\geq \log_2(m)-\log(c)-\eta \phi -\log(\eta p^{*})-m\\
    &-\log(2\pi e)-\log(k^{'})+\log(12)+\max_{i}h(\text{Poisson}(M \tilde{p_{i}})).
\end{align*}
where term $h(\text{Poisson}(M p_{j}^{'})$ is computed with results of \cite{8437888} as stated in \eqref{eq01-1}.
\newline
Now, we consider the case of large $m$. Noting  $\log(\frac{x}{m})\leq \frac{x}{m}-1$, a lower bound on $h(Y_i)$ is
\begin{equation}
    h(Y_i)\geq h(X,T_x)-\log(\eta p_{i}^{*})-\log(m)-1-\log(k^{'}).
\end{equation}
Thus, we have
\begin{align*}
   I(T_x,X;Y_1,...,Y_n)\geq \log_2(m)-\log(c)-\eta \phi -\log(\eta p^{*})-\log(m)\\
    -1-\log(2\pi e)-\log(k^{'}).
 +\log(12)+\max_{i}h(\text{Poisson}(M \tilde{p_{i}})),
\end{align*}
which reduces to \eqref{eq01-2} for large $m$, if the average constraint in (1) holds with equality.
\section{Proof of \eqref{diff}}
 To lower bound $I(T_x;Y_i-Y_{i-1})$, we write
\begin{equation}
I(T_x,Y_i-Y_{i-1}|X)=h(Y_i-Y_{i-1}|X)-h(Y_i-Y_{i-1}|X,T_x).\label{here22}
\end{equation}
Using \eqref{here44}, we have:
\begin{flalign}
\!\!\!\!\!\!\!Pr(Y=y_i-y_{i-1}|X=x,T_x=t_x)=\frac{e^{\frac{-1}{2x q_{ij}}( y_i-x(q_{ij}))^{2}}}{\sqrt{2\pi x(q^{'}_{ij})}},\label{here59}
\end{flalign}
where $q_{ij}=p_{ij}-p_{i-1j}$ and $q^{'}_{ij}=p_{ij}+p_{i-1 j}$. Thus,
\begin{flalign}
&Pr(Y_i-Y_{i-1}=y|X=x)=\sum_{j=1}^{m}Pr(T_x=j\sigma_x) \frac{e^{\frac{-( y-x(q_{ij}))^2}{2x q^{'}_{ij}}}}{\sqrt{2\pi x(q^{'}_{ij}))}}.\label{here61}
\end{flalign}
\begin{lemma}
The entropy of mixture of Gaussian in \eqref{here61} is lower bounded using Moments of its Gaussian components.
\end{lemma}
\begin{proof}
Proof is based on using Taylor series representation of logarithm of Gaussian Mixture and also using differential entropy. We use a similar approach of \cite{huber2008entropy} for scalar Gaussian random variables. without loss of generality, we assume $E\{X\}=\eta$. For simplicity, we define the following notation for \eqref{here61}.
\begin{equation*}
  g(y)\overset{\triangle}{=} Pr(Y_i-Y_{i-1}=y|X=x)= \sum_{j}^{}a_{ij} e^{b_{ij}(y-c_{ij})^2}
\end{equation*}
where $a_{ij}$, $b{ij}$ and $c_{ij}$ are evident from \eqref{here61}. Thus,
 \begin{equation*}
      \log(g(y))=\sum_{k=0}^{n} z_k (y-y_{i0})^{k}+R_n,
\end{equation*}
where $y_{i0}=xq_{i1}$ and $z_k$ is the coefficients of Taylor series for $\log(g(y))$, and $y_{i0}=xq_{i1}$.
Therefore:
\begin{flalign}
  &h(Y_i-Y_{i-1}|X)=\int_{y}^{} \sum_{j}^{}a_{ij} e^{b_{ij}(y-c_{ij})^2}\log( \sum_{j}^{}a_{ij} e^{b_{ij}(y-c_{ij})^2}) dy, \notag\\
   &\overset{(a)}{\geq}\int_{y}^{} \sum_{j}^{}a_{ij} e^{b_{ij}(y-c_{ij})^2}(\sum_{k=0}^{r}z_k(y-y_0)^{k}) dy \notag\\
   &\overset{(b)}{\geq}\sum_{k=0}^{r} \sum_{j}^{}a_{ij}z_k m_{k i j}, \label{here1000}
\end{flalign}
where (a) follows by approximating the logarithm with first $r$ terms in Taylor series and (b) follows by using convexity of $e^x$, Jensen's inequality, and noting that $m_{kij}$ is the non-central moments of a Gaussian random variable with mean $ \eta q_{il} $ and variance $\eta q^{'}_{il}$, computed using its central moments $\mu_{il}$ as
\begin{equation}\label{here1001}
    m_{kij}=\sum_{l=0}^{k}\binom{k}{l}\mu_{il} ^{l}(c_{ij}-y_0)^{n-l}.
\end{equation}
which completes our proof.
\end{proof}
Now, we consider the second term in \eqref{here22}.
\begin{lemma}
The entropy of probability distribution in \eqref{here59} is:
\begin{equation*}
 \!\!h(Y_i-Y_{i-1}|X,T_x)\!=\!\!\sum_{j}^{}Pr(T_x=j\sigma_x)\frac{1}{2}\log(2 \pi e  q^{'}_{ij} +\frac{1}{2} E\{\log X\}).
 \end{equation*}
 \end{lemma}
\begin{proof}
  We have:
\begin{flalign*}
& h(Y_i-Y_{i-1}|X,T_x)= \sum_{j}^{}Pr(T_x=t_x)h(Y_i-Y_{i-1}|X,T_x=j\sigma_x)\\
&\overset{(a)}{=}\sum_{j}^{}Pr(T_x=t_x)(\frac{1}{2}\log(2 \pi e  q^{'}_{ij}+\frac{1}{2} E\{\log X\})).
\end{flalign*}
where (a) follows from the Gaussian entropy.
\end{proof}
By using Lemmas 5 and 6, we conclude that:
\begin{align}
     I(T_x;Y_i&-Y_{i-1}|X)\geq \sum_{k=0}^{r} \sum_{j}^{}a_{ij}z_k m_{k i j}\label{here1002}\\
    &+ \sum_{j}^{}Pr(T_x=j\sigma_x)(\frac{1}{2}\log(2 \pi e  q^{'}_{ij}+\frac{1}{2} E\{\log X\})).\notag
\end{align}
 We propose using distribution $f_{X,T_x}(x,j\sigma_x)=\frac{1}{\sqrt{\pi u} m}e^{\frac{-x^2}{u}}$ for computing the above lower bound (where $u$ is chosen such that the distribution integrates to one), which results in:
\begin{align*}
     I(T_x;Y_i-Y_{i-1}&|X)\geq\frac{1}{4 m}(\log(M)\erf{\frac{M}{\sqrt{u}}})\\
       &-\frac{1}{2 m \sqrt{\pi u}}M \Hypergeometric{2}{2}{\frac{1}{2},\frac{1}{2}}{\frac{3}{2},\frac{3}{2}}{\frac{-M^2}{u}}\\
     &\sum_{k=0}^{r} \sum_{j}^{}a_{ij}z_k m_{k i j}+\frac{1}{2 m} \sum_{j}^{}\frac{1}{2}\log(2 \pi e  q^{'}_{ij}).
\end{align*}
Combining this bound with the lower bound in \eqref{eq3-3} (with the above proposed distribution $f_{X,T_x}(x,j\sigma_x)$) and considering constraint (1) hold with equality, we obtain \eqref{diff}.
\end{document}